\documentclass[prd,preprint,nofootinbib]{revtex4}
\usepackage{amssymb}
\usepackage{epsfig}
\newcommand{\bea}{\begin{eqnarray}}  \newcommand{\eea}{\end{eqnarray}}
\newcommand{\beq}{\begin{equation}}  \newcommand{\eeq}{\end{equation}}
\newcommand{\bef}{\begin{figure}}  \newcommand{\eef}{\end{figure}}
\newcommand{\bec}{\begin{center}}  \newcommand{\eec}{\end{center}}
\newcommand{\non}{\nonumber}  
\newcommand{\lmk}{\left(}  \newcommand{\rmk}{\right)}
\newcommand{\lkk}{\left[}  \newcommand{\rkk}{\right]}
  
\newcommand{\lnk}{\left \{ }  \newcommand{\rnk}{\right \} }

\newcommand{\phip}{\phi_{+}}
\newcommand{\phim}{\phi_{-}}

\newcommand{\re}{\mbox{Re}}

\newcommand{\fp}{f_+}
\newcommand{\fm}{f_-}
\newcommand{\fps}{f_{+\sigma}}
\newcommand{\fms}{f_{-\sigma}}
\newcommand{\fpms}{f_{\pm\sigma}}
\newcommand{\fpm}{f_\pm}
\newcommand{\cp}{c_+}
\newcommand{\cm}{c_-}
\newcommand{\cpm}{c_\pm}
\newcommand{\phipm}{\phi_\pm}
\begin{document}
\title{ 
Hiding cosmic strings in supergravity D-term inflation
}

\author{Osamu Seto}
\affiliation{
 Department of Physics and Astronomy, University of Sussex, 
 Brighton BN1 9QJ, United Kingdom}

\author{Jun'ichi Yokoyama}
\affiliation{
 Research Center for the Early Universe, School of Science,
The University of Tokyo, Tokyo 113-0033, Japan}

%
\begin{abstract}
The influence of higher-order terms in the K\"{a}hler potential 
of the supergravity D-term inflation model
 on the density perturbation is studied.  We show that these terms
can make the inflaton potential flatter, which lowers the energy 
scale of inflation under the COBE/WMAP normalization.
As a result, the mass per unit length of cosmic strings, which
are produced at the end of inflation, can be reduced
 to a harmless but detectable level without introducing a
tiny Yukawa coupling.  Our scenario can naturally be implemented
in models with a low cut-off as in 
Type I or Type IIB orientifold models.

\end{abstract}

\preprint{RESCEU-35/05 } 

\vspace*{3cm}
\maketitle


\section{Introduction}

Inflation in the early Universe not only realizes globally 
homogeneous and flat space but also provides the seeds of
density perturbations \cite{Inflation}.
To realize successful inflation which matches observational
data of large-scale structures and anisotropy of cosmic microwave
background radiation (CMB), the potential of the scalar field which
drives inflation, {\it the inflaton}, 
 must be very flat.
The required flat potential may be realized with the help of
supersymmetry or supergravity. 
In supersymmetric models, the scalar potential consists of
 the contribution from F-term and D-term.
In F-term inflation models where the vacuum energy 
to drive  inflationary
 expansion is dominantly provided by the F-term,
 the inflaton mass is in general of 
the same order of the Hubble parameter $H$
 during inflation, in other words, a slow roll parameter
\begin{equation}
\eta \equiv \frac{V''[\sigma]}{V[\sigma]} ,
\end{equation}
generally takes
 a value of the order of unity, although $\eta\ll 1$ should be satisfied
 for  successful inflation.
 Here $V[\sigma]$ is the potential energy density of the inflaton,
 $\sigma,$ 
 the prime denotes the derivative with respect to the inflaton field and
 we take the unit with $8\pi G=1$ throughout this paper. 
It is therefore difficult to realize a sufficiently  long expansion
 to solve the horizon and flatness problems.
This is the 
so-called the $\eta$ problem of inflation models in supergravity.
On the other hand, D-term inflation,
 where the vacuum energy is provided by D-term,
 does not suffer from the problem
 \cite{DtermInflation}.
Hence, D-term inflation appears more attractive than F-term inflation
 from this point of view.

However, it has been revealed that the D-term inflation model also has
 some problems \cite{LythRiotto}.
For example, from the observational point of view,
 the cosmic strings generated after inflation significantly affect
 the spectrum of CMB anisotropy
 \cite{Jeannerot}, 
 because this is a kind of hybrid inflation model \cite{Hybrid}.
In addition, one may suspect the potential for D-term inflation 
not to be valid,
 because the inflaton needs to have a large initial value of the order of
 (sub-)Planck scale for a natural model parameter \cite{Kolda}.
Hence, D-term inflation seems to be under strong pressure.
In addition, in the framework of (heterotic) string models, 
there are two more
 problems, namely,
 the runaway behavior of dilaton and too large a magnitude of
 Fayet-Illiopoulos (FI) term. (See, however, \cite{King}.)

In this paper, we investigate effects of higher-order terms 
in the K\"{a}hler
 potential in D-term inflation.
So far,  little attention has been paid to the effects of
 the K\"{a}hler potential in the dynamics of D-term inflation,
 because these terms do not disturb the flatness of the potential
 unlike in F-term inflation models.
However, we should clarify the effects of higher-order terms
in the K\"{a}hler potential,
 since the inflaton must take  a large initial value close to the
Planck scale.
We study the effects of higher-order terms in the K\"{a}hler potential,
unlike Rocher and Sakellariadou who 
studied D-term inflation recently  taking only 
 the leading term in the K\"{a}hler potential into account \cite{Rocher}.

As we will show, these terms alter the dynamics of the inflation and
 the resultant constraints on the model parameter,
 especially the magnitude of FI term.
Hence, as the result,
 the predicted mass per unit length of cosmic strings can be reduced
 and meet the observational constraints without assuming a very small 
Yukawa
 coupling as in previous works \cite{Rocher, Endo}.

The rest of this paper is organized as follows.
After reviewing the D-term inflation model in the minimal supergravity
 in the next section, in Secs. III and IV, we study the effects of
two types of higher-order terms separately.
Section V is devoted to conclusions.

\section{D-term inflation}

Here we give a brief review on the D-term inflation
 \cite{LythRiotto,DtermInflation} in minimal supergravity 
 in order to make the problem clear and help to compare it with our model
 in the following section.
We consider the $\mathcal{N}=1$ supersymmetric model with $U(1)$ 
gauge group and
 the non-vanishing FI term $\xi$.
The minimal model contains three matter fields $S$ and $\phi_\pm$.
The fields $\phi_\pm$ have $U(1)$ charges $q_{\pm}=\pm 1$ such 
that $\xi > 0$, while $S$ is neutral for the $U(1)$.
Suppose the following K\"ahler potential and superpotential,
\bea
K &=& |S|^2+|\phip|^2+|\phim|^2, \\
W &=& \lambda S \phi_+ \phi_- .\label{Superpotential}
\eea
Then, the scalar potential is written as
\bea
V &=& \lambda^2 e^{|S|^2+|\phip|^2+|\phim|^2}
\lkk |\phip\phim|^2+|S\phim|^2+|S\phip|^2 
+(|S|^2+|\phip|^2+|\phim|^2+3)|S\phip\phim|^2\rkk \non\\
&&~~+ 
\frac{g^2}{2}\left(\xi + |\phi_+|^2 - |\phi_-|^2\right)^2 ,
\eea
where $g$ is the gauge coupling.
The true vacuum of this potential corresponds to
\begin{equation}
S= \phi_+ =0, \qquad |\phi_-| = \sqrt{\xi} .
\end{equation}

For a large value of $S$,
\begin{equation}
|S|>S_c \equiv \frac{g}{\lambda} \sqrt{\xi},
\end{equation}
the potential has the local minimum with a non-vanishing vacuum energy
density
\begin{equation}
V_0 = \frac{g^2}{2} \xi^2
\end{equation}
at
\begin{equation}
|\phi_\pm| =0 ,
\end{equation}
which minimizes the potential in this regime.
Then, the radial part of $S$ is a flat direction but it acquires
a non-vanishing potential through radiative corrections, for
supersymmetry is broken due to the non-vanishing D-term.
In this regime the scalar fields $\phi_\pm$ have masses
\begin{equation}
m^2_\pm = \lambda^2 |S|^2e^{|S|^2} \pm g^2 \xi,
\end{equation}
while the mass-squared of their fermionic partner is simply given 
by $\lambda^2 |S|^2e^{|S|^2}$.
As a result the one-loop effective potential is given as
\begin{equation}
V_{1-loop} = \frac{g^2}{2} \xi^2 \left(1
+ \frac{g^2}{8 \pi^2}
\ln {\frac{\lambda^2 |S|^2e^{|S|^2}}{\Lambda^2} } \right) ,
\end{equation}
for a large field value $|S|^2 \gg g^2\xi/\lambda$,
where $\Lambda$ is a renormalization scale.

Without loss of generality we can identify the real part of $S$,
$\sigma\equiv \sqrt{2}\re S$, as
the inflaton.
Thus, the inflaton slowly rolls down the potential from a large 
initial value during inflation.
When the inflaton reaches
\begin{equation}
\sigma_c \equiv \sqrt{2}S_c = \frac{\sqrt{2\xi}g}{\lambda}, \label{S_c}
\end{equation}
where $\phi_-$  becomes tachyonic or
\begin{equation}
\sigma_f \equiv \frac{g}{2\pi},
\end{equation}
which corresponds to $\eta=V''[\sigma]/V_0=-1$, the inflation terminates.
Unless the Yukawa coupling $\lambda$ is extremely small, 
$\lambda \lesssim 10^{-4}$,
 inflation terminates when the inflaton arrives at $\sigma_f$.
In the late stage of inflation when $e^{\sigma^2/2}\approx 1$, 
the inflaton evolves as
\begin{equation}
\frac{\sigma^2}{2}-\frac{\sigma_e^2}{2} = \frac{g^2}{4\pi^2}N, 
\label{PhiSolution}
\end{equation}
where $\sigma_e = \max(\sigma_f,\sigma_c)$ is the field value at the 
end of inflation and
$N$ is the number of e-folds acquired between $\sigma$ and $\sigma_e$. 
For $N= 50-60$ and a natural value of gauge coupling $g$,
 the right-hand side of
 Eq.\ (\ref{PhiSolution}) is $\mathcal{O}(0.1)-\mathcal{O}(1)$.
This means that the inflaton must take a large field value
 of the order of sub-Planckian scale.

In the case  $\sigma_e = \sigma_f$, by using Eq. (\ref{PhiSolution}),
the amplitude of the comoving curvature perturbation is given as
\begin{eqnarray}
\mathcal{P}^{1/2}_{\zeta} \equiv \frac{H^2}{2\pi|\dot{\sigma}|}
 &=& \xi\sqrt{\frac{N}{3}} \label{SimpleAmplitude} \nonumber \\
 &=& 4.7\times 10^{-5}\left(\frac{\xi}{1.1 \times 10^{-5}}\right)\left(\frac{N}{55}\right)^{1/2}.
\end{eqnarray}
Thus, the required magnitude of FI term $\xi$ to generate the appropriate
 amplitude of the density perturbation is estimated as
 $\xi=1.1 \times 10^{-5}$, in other words,
\begin{equation}
 \sqrt{\xi} \simeq 3.3 \times 10^{-3} \simeq 7.6 \times 10^{15} \textrm{GeV}.
 \label{InflationFI}
\end{equation}
After inflation,
 the cosmic string with the mass per unit length $2\pi\xi$ is generally formed,
 although an exceptional model has been proposed \cite{Urrestilla}.
These cosmic strings potentially affect the density perturbation, indeed,
 this fact is a fatal shortcoming for D-term inflation model \cite{Jeannerot}.
The constraints on cosmic strings have been studied \cite{Endo, Other}.
According to a recent study \cite{Endo},
 the constraints on the magnitude of the FI term is derived as
\begin{equation}
\sqrt{\xi} \lesssim 1.9 \times 10^{-3}, \label{StringConstraint}
\end{equation}
and obviously conflicts with Eq.(\ref{InflationFI}).

It has been proposed that this problem might be avoided by considering
 the other case $\sigma_e = \sigma_c > \sigma_f$ with a very small
Yukawa coupling,  $\lambda \lesssim 10^{-4}$
 \cite{Endo}.
This, however,
 does not entirely solve the problem, because taking a small 
$\lambda$ results in a larger value of $S_c$. If its value exceeds
unity,
the supergravity effect or the exponential factor in
the potential plays an important role.  Then the above formula
of curvature perturbation can no longer be used and we would
find a blue-tilted
spectrum.  As a result, we again need a larger value of $\xi$ to
generate the density perturbation with the appropriate magnitude.
Indeed, the only way to escape from this difficulty is to adopt an 
anomalously small value of $g \lesssim 10^{-2}$ \cite{Rocher},
as is seen from the relation
\begin{equation}
S_c = \frac{\sqrt{\xi}g}{\lambda} 
= 1 \left(\frac{\sqrt{\xi}}{10^{-3}}\right)\left(\frac{10^{-5}}{\lambda}\right)
\left(\frac{g}{10^{-2}}\right) \lesssim 1.
\end{equation}

\section{D-term inflation with higher-order coupling between
the inflaton and the charged fields}

In this section, we take higher-order terms in the K\"{a}hler potential
 into account in analyses of the dynamics of D-term inflation.
Since we focus on the inflation regime when $\phi_+$ and $\phi_-$
are small, the following K\"ahler potential can be regarded as a
sufficiently  general one.
\begin{equation}
K = |S|^2+|\phi_+|^2+|\phi_-|^2+\fp(|S|^2)|\phi_+|^2
+\fm(|S|^2)|\phi_+|^2 ,
\end{equation}
where $f_\pm(|S|^2)$ are arbitrary functions of $|S|^2$.
One can also add higher-order terms containing $S$ alone, but this
can be absorbed by a field redefinition of $S$. 
Here  we do not
consider such terms and assume that $S$ has a canonical kinetic
term at $\phi_\pm=0$.
The effect of such a redefinition will be studied separately
in the next section.

With the superpotential Eq. (\ref{Superpotential}),
 the Lagrangian density is given by
\begin{equation}
{\cal L} = {\cal L}_{kin} - V
\end{equation}
with
\begin{eqnarray}
{\cal L}_{kin}
&=& -(\partial S, \partial\phi_+, \partial\phi_-) \nonumber \\
&\times& \!\!
\left(
\begin{array}{ccc}
1+(\fp'+\fp''|S|^2)|\phi_+|^2
 +(\fm'+\fm''|S|^2)|\phi_-|^2 & \fp'S^*\phi_+ & \fm'S^*\phi_+ \\
\fp'\phi_+^* & 1+\fp & 0 \\
\fm'S\phi_-^* & 0 & 1+\fm
\end{array}
\right)\!
\left(
\begin{array}{c}
\partial S^* \\
\partial \phi_+^* \\
\partial \phi_-^*
\end{array}
\right)
\end{eqnarray}
and
\begin{eqnarray}
&&V = V_F + V_D, \\
&&V_F = \lambda^2 \frac{e^K }{\det K_i{}^j}
\biggl[
\lnk (1+\fp)(1+\fm)-\lkk (1+\fp)(1+\fm)\rkk'|S|^2\rnk|\phip\phim|^2
\nonumber \\
&&+\lnk (1+\fm)(1+\fp''|S|^2|\phip|^2)
+ \lkk (1+\fm)(\fm'+\fm''|S|^2) -\fm'^2|S|^2\rkk|\phim|^2
+\fp'\fm'|S\phim|^2 \rnk |S\phim|^2 \nonumber \\
&&+\lnk (1+\fp)(1+\fm''|S|^2|\phim|^2)
+ \lkk (1+\fp)(\fp'+\fp''|S|^2) -\fp'^2|S|^2\rkk|\phip|^2
+\fp'\fm'|S\phip|^2 \rnk |S\phip|^2
\biggl] \nonumber\\
&&-3\lambda^2 e^K|S\phip\phim|^2, \\
&&V_D = \frac{g^2}{2}\left(q_+(1+\fp)|\phi_+|^2
+q_-(1+\fm)|\phi_-|^2+\xi\right)^2,
\end{eqnarray}
where
\beq
  f'_\pm\equiv \left.\frac{d f_\pm(x)}{dx}\right|_{x=|S|^2},~~
  f''_\pm \equiv \left.\frac{d^2 f_\pm(x)}{dx^2}\right|_{x=|S|^2}.
\eeq
Here the determinant of the K\"{a}hler metric is given by
\begin{eqnarray}
\det K_i{}^j &=& \lkk 1+(\fp'+\fp''|S|^2)|\phi_+|^2
+(\fm'+\fm''|S|^2)|\phi_-|^2\rkk(1+\fp)(1+\fm)  \nonumber\\
&& -(1+\fp)\fm'^2|S\phi_-|^2 - (1+\fm)\fp'^2|S\phi_+|^2.
\end{eqnarray}

Here and hereafter, we assume that both
$\fp$ and $\fm$ are well behaved in the sense
 that the kinetic terms have the correct signature.

Equations of motion of $\phi_{\pm}$ are given by
\begin{equation}
-(1+\fpm)\partial^2\phi_{\pm} -2\fpm' S^*\partial S \partial\phi_{\pm}
-\fpm'S^*\phipm\partial^2S -\fpm''\phipm(S^*\partial S)^2
+\frac{\partial V}{\partial \phi_{\pm}^*}=0 .
\end{equation}
The second, third, and fourth terms are
additional terms through the higher-order terms
 and regarded as the additional friction term
 and the additional mass terms, respectively.
Since both are negligible during inflation by the slow
roll conditions
 $|\dot{S}/S|\ll H$ and $|\ddot{S}/\dot{S}|\ll H$,
 they do not affect the dynamics of inflation.
Then, the equation of motion of $\phi_+$ is given as
\begin{equation}
(1+\fp)\left(\ddot{\phi_+}+3H\dot{\phi_+}+q_+g^2\xi\phi_+\right)+
e^{|S|^2}\frac{\lambda^2|S|^2}{(1+\fm)}\phi_+\simeq 0,
\end{equation}
near $\phip=0$, and $\phi_+$ settles at $\phi_+ =0$.
Similarly, the equation of motion of $\phi_-$ is given as
\begin{equation}
(1+\fm)\left(\ddot{\phi_-}+3H\dot{\phi_-}+q_-g^2\xi\phi_-\right)
+e^{|S|^2}\frac{\lambda^2|S|^2}{(1+\fp)}\phi_-\simeq 0.
\end{equation}
Now the critical point $S_c$ at which $\phim=0$ becomes unstable is
given by a solution of
\beq
  e^{|S|^2}\frac{\lambda^2|S|^2}{(1+\fp)(1+\fm)}=-q_-g^2\xi.
\eeq
If we assume $\fpm$ are small for small $S$, $S_c$
is practically the same as Eq. (\ref{S_c}) for $\lambda = \mathcal{O}(1)$
 because $\xi \ll 1$.
Thus, we find that inflation would successfully
proceed in the same manner as
 the simple model which we have reviewed in the previous section.

Next, we include the radiative correction by $\phi_{\pm}$ and derive the
 one-loop effective potential.
Here, we should notice that $\phi_{\pm}$ are not canonical
any longer owing to
 the mixing terms in the K\"{a}hler potential.
While the effective masses of charged scalar fields $\phi_{\pm}$
is given as
\begin{equation}
m_{\varphi_{\pm}}^2 = q_{\pm}g^2\xi+e^{|S|^2}
\frac{\lambda^2|S|^2}{(1+\fp)(1+\fm)}
\label{mphi}
\end{equation}
 for the canonically normalized variables $\varphi_{\pm}$,
 the mass of their fermionic superpartners is given as
\begin{equation}
m_{fermion}^2 = e^{|S|^2}\frac{\lambda^2|S|^2}{(1+\fp)(1+\fm)}
\label{mfer}
\end{equation}
 for the canonical variable.
The correction from one-loop contributions to the potential
is expressed as
\begin{eqnarray}
\delta V
&=& \frac{1}{32\pi^2}\left(q_+g^2\xi+e^{|S|^2}
\frac{\lambda^2|S|^2}{(1+\fp)(1+\fm)}\right)^2\ln\left(\frac{q_+g^2\xi+e^{|S|^2}
\frac{\lambda^2|S|^2}{(1+\fp)(1+\fm)}}{\Lambda^2}\right) \nonumber \\
&& +\frac{1}{32\pi^2}\left(q_-g^2\xi+e^{|S|^2}
\frac{\lambda^2|S|^2}{(1+\fp)(1+\fm)}\right)^2\ln\left(\frac{q_-g^2\xi+e^{|S|^2}
\frac{\lambda^2|S|^2}{(1+\fp)(1+\fm)}}{\Lambda^2}\right) \nonumber \\
&& -\frac{1}{16\pi^2}\left(e^{|S|^2}
\frac{\lambda^2|S|^2}{(1+\fp)(1+\fm)}\right)^2\ln\left(\frac{e^{|S|^2}
\frac{\lambda^2|S|^2}{(1+\fp)(1+\fm)}}{\Lambda^2}\right) \nonumber \\
&\simeq& \frac{1}{16\pi^2}g^4\xi^2
\left[ \ln\left(\frac{\lambda^2|S|^2}{(1+\fp)(1+\fm)\Lambda^2}\right) +|S|^2\right]  \qquad \textrm{for \, large}\,\, S.
\end{eqnarray}
We can identify the real part of $S$,
 $\sigma=\sqrt{2}{\rm Re}S$, with the
inflaton as before and we obtain the one-loop effective potential
\begin{equation}
V_{1-loop}(\sigma)=\frac{g^2\xi^2}{2}\left(1+\frac{g^2}{8\pi^2}
\left[\ln\frac{\lambda^2\sigma^2}
{(1+\fp)(1+\fm)\Lambda^2}
+\frac{\sigma^2}{2}\right]\right) , \label{Potential}
\end{equation}
with $\fpm=\fpm(\sigma^2/2)$
for a large field value of $\sigma$.
The last term comes from the exponential
 factor of the mass of charged fields. 

Field equations during inflation with $\phipm=0$ read
\begin{equation}
H^2 = \frac{1}{3}\frac{g^2\xi^2}{2} ,
\end{equation}
and
\begin{equation}
\ddot{\sigma} +3H\dot{\sigma}+ \frac{g^2\xi^2}{2}\frac{g^2}{8\pi^2}
\left(\frac{2}{\sigma}-\frac{\fps}{1+\fp}
-\frac{\fms}{1+\fm}+\sigma\right)= 0 .\label{EOM}
\end{equation}
The additional new terms from the higher coupling
 between the inflaton and charged fields in
 the K\"{a}hler potential can be important when
$\fpms/(1+\fpm)$ are comparable to $2/\sigma$ or $\sigma$.
Here the suffix $\sigma$ denotes differentiation with respect to
$\sigma$.
In order for $\sigma$ to evolve towards $\sigma_e$,
\beq
  V'[\sigma]=  \frac{g^2\xi^2}{2}\frac{g^2}{8\pi^2}
\left(\frac{2}{\sigma}-\frac{\fps}{1+\fp}
-\frac{\fms}{1+\fm}+\sigma\right),
\eeq
should be  positive definite
\footnote{
However, if the local maximum is located
 on a larger field value $\sigma > \sigma_{N\sim 55}$,
 it might not be necessary for this requirement to be strict.
 In fact, the eternal inflation scenario could be realized
 around the very flat hill \cite{Linde}.
 One may find a similar discussion in Ref. \cite{Hilltop}.
 }.
The amplitude of the comoving curvature perturbation in this case
reads
\begin{eqnarray}
\mathcal{P}^{1/2}_{\zeta}&=&\frac{H^2}{2\pi|\dot{\sigma}|}
=\frac{g^3\xi^3}{4\sqrt{6}\pi|V'[\sigma]|} \nonumber\\
&=&\frac{4\pi\xi}{\sqrt{6}g}
\lmk \frac{2}{\sigma}+\sigma -\frac{\fps}{1+\fp}-\frac{\fms}{1+\fm}
\rmk^{-1}, \label{yuragi}
\end{eqnarray}
under the slow-roll approximation.

First let us incorporate lowest order correction to the K\"ahler
potential,
\beq
  \fpm=\frac{\cpm}{2}\sigma^2,
\eeq
with $\cpm$ being positive constants.  If we take $\cp=\cm\equiv c$,
the  positivity
 condition $V'[\sigma]> 0$ requires $c< 3+2\sqrt{2}$.
Conversely, if we take $c=3+2\sqrt{2}\simeq 5.83$, $V'[\sigma]$ vanishes at
$\sigma= \sqrt{2(\sqrt{2}-1)}\simeq 0.91$. 
This means that if we
take $c$ slightly smaller than the critical value $3+2\sqrt{2}$ the
potential is very flat near $\sigma\simeq 0.91$. 
Then the amplitude of density
fluctuation is enhanced due to this flatness of the potential and
we can achieve the desired amplitude, $\approx 10^{-5}$ with smaller
values of $\xi$.  From (\ref{yuragi}) we find that if we take
$c=5.5$ and $g=0.7$, the amplitude of curvature fluctuation meets
the CMB normalization with a small enough value of $\xi$,
$\xi=2.7\times 10^{-6}$, as is seen in Figs.\ 1 and 2, where
the spectral index takes $n_s \simeq 0.96.$

\begin{figure}
\epsfxsize=0.7\textwidth
\centerline{\epsfbox{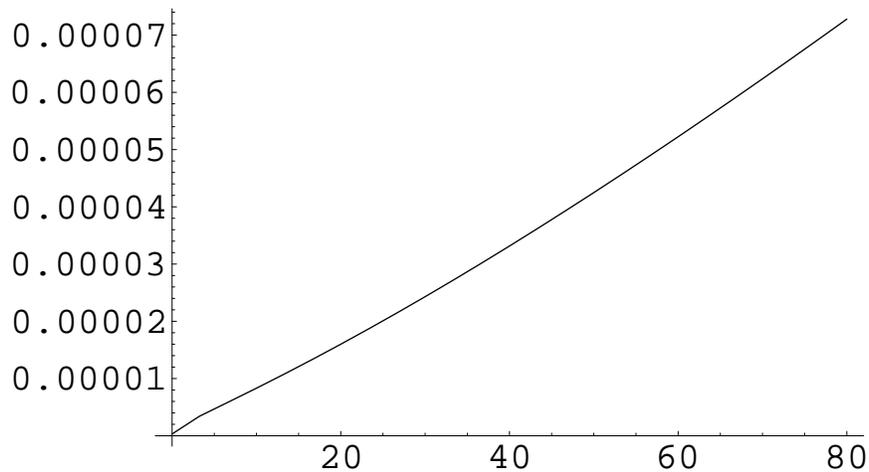}}
\caption{
The amplitude of density perturbation
 for parameters $g=0.7,\, c_+ = c_-= 5.5$ and
  $\xi= 2.7 \times10^{-6}$.
A horizontal axis represents the number of e-folds $N$ and a
vertical axis
 represents the amplitude of density
perturbation $\mathcal{P}^{1/2}_{\zeta}$.
$N \sim 55$ would correspond to the present horizon scale.
 }
\end{figure}

\begin{figure}
\epsfxsize=0.7\textwidth
\centerline{\epsfbox{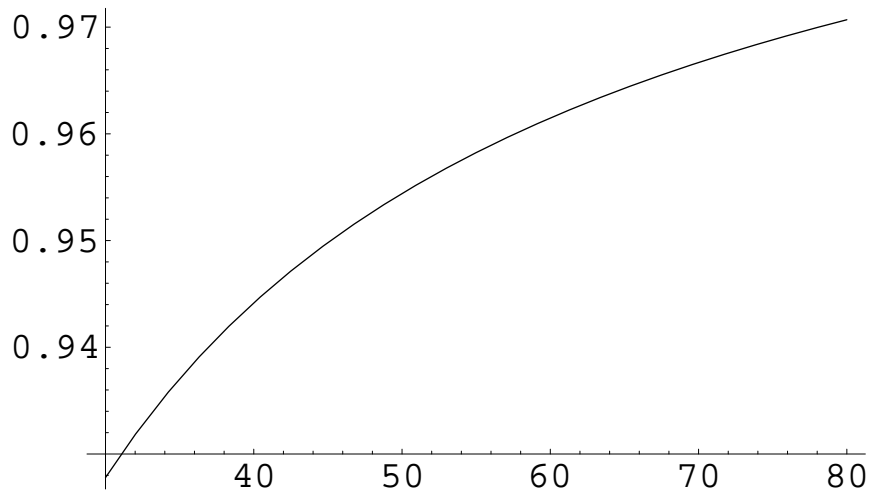}}
\caption{
The spectral index for the same parameters in Fig.1.
A horizontal axis represents the number of e-folds $N$ and
 a vertical axis represents the spectral index $n_s$.
 }
\end{figure}

Next we consider the case $c_+ \neq c_-$.  Without loss of generality
we can assume that $c_+ > c_-$.  In this case the potential of $\sigma$
can remain monotonic even if $\cp$ is larger than $3+2\sqrt{2}$.
In particular, if we take $\cm=0$, $\cp$ can be arbitrary large.

For small $\sigma$, $\sigma \lesssim \sqrt{2/c_+}$,
 the solution of the equation
of motion is the same as Eq. (\ref{PhiSolution})
 with $\sigma_e = \sigma_f$,
since we consider the case of $\lambda=\mathcal{O}(1)$.
On the other hand, for a region of a large field value,
 $\sigma\gtrsim \sqrt{2/c_+}$,
 the effect from the coupling in the K\"{a}hler potential becomes
significant.
For $\sigma\gg \sqrt{2/c_+}$, the
slow-roll equation (\ref{EOM}) is rewritten as
\begin{equation}
3H\dot{\sigma}
+\frac{g^2\xi^2}{2}\frac{g^2}{8\pi^2}\left(\frac{2}{\frac{c_+}{2}\sigma^3}
-\frac{c_-\sigma}{1+\frac{c_-}{2}\sigma^2}+\sigma\right)\simeq 0 .
\end{equation}
Here, as we expect, the additional
 term with $c_+$ makes the potential flatter by
 cancelling out the leading term $2/\sigma$ in $V'[\sigma]$.
The approximate solution is given as
\begin{equation}
\frac{\sigma^4}{4}-\frac{\sigma^4_*}{4} = \frac{g^2}{4\pi^2}\frac{2}{c_+}(N-N_*),
\end{equation}
for the region where the first term dominates other terms in
$V'[\sigma]$,
where $\sigma_* \simeq \sqrt{2/c_+}$ and the corresponding number
of e-folds
 $N_* \simeq 4\pi^2/(g^2 c_+)$.

We turn to the generated density perturbation.
Here we consider the case that the present horizon scale corresponds to
 a large field value of $\sigma$, $\sigma^2 \gg 2/c_+$.
Then the amplitude of the density perturbation is estimated as
\begin{eqnarray}
\mathcal{P}_{\zeta}^{1/2} &=&
 \frac{\sqrt{\frac{g^2\xi^2}{6}}}{2\pi\frac{g^2}{4\pi^2}\frac{2}{c_+\sigma^3}}
\left(1-\frac{c_-}{1+\frac{c_-}{2}\sigma^2}\frac{c_+\sigma^4}{4}
+\frac{c_+\sigma^4}{4}\right)^{-1} \\
&\simeq& \xi\sqrt{\frac{N}{3}}\left(\frac{4g^2}{\pi^2}\frac{c_+}{2}N\right)^{1/4}
 \label{Amplitude} \\
&\simeq& 4.5\times 10^{-5} \left(\frac{\xi}{3.6 \times 10^{-6}}\right)
\left(\frac{N}{55}\right)^{1/2+1/4}\left(\frac{g^2}{0.8}\right)^{1/4}
\left(\frac{c_+}{8}\right)^{1/4} \label{NormalizedAmplitude} ,
\end{eqnarray}
where we omitted the $(...)^{-1}$ factor after Eq. (\ref{Amplitude})
 for simplicity.
Comparing Eq. (\ref{Amplitude}) with Eq. (\ref{SimpleAmplitude}),
we find that,
 if $(2g^2c_+N/\pi^2)$ is greater than unity, the factor enhances the amplitude of the density perturbation.
Indeed, as is shown in Eq. (\ref{NormalizedAmplitude}),
 it is possible for natural values of parameters.
Then, the magnitude of FI term is significantly reduced and satisfies
 the condition Eq. (\ref{StringConstraint}) from the cosmic string constraint.
The results of more exact numerical analysis are shown in Figs.\ 3 and
 4.

\begin{figure}
\epsfxsize=0.7\textwidth
\centerline{\epsfbox{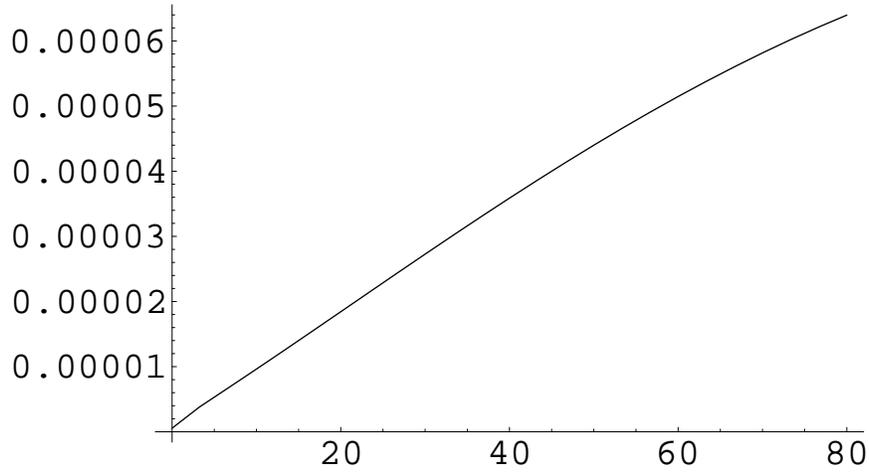}}
\caption{
The amplitude of density perturbation
 for parameters $g=0.9,\, c_+ = 8,\, c_-=3$ and
  $\xi= 2.7 \times10^{-6}$.
A horizontal axis represents the number of e-folds $N$ and a
vertical axis
 represents the amplitude of density
perturbation $\mathcal{P}^{1/2}_{\zeta}$.
$N \sim 55$ would  correspond to the present horizon scale.
 }
\end{figure}

\begin{figure}
\epsfxsize=0.7\textwidth
\centerline{\epsfbox{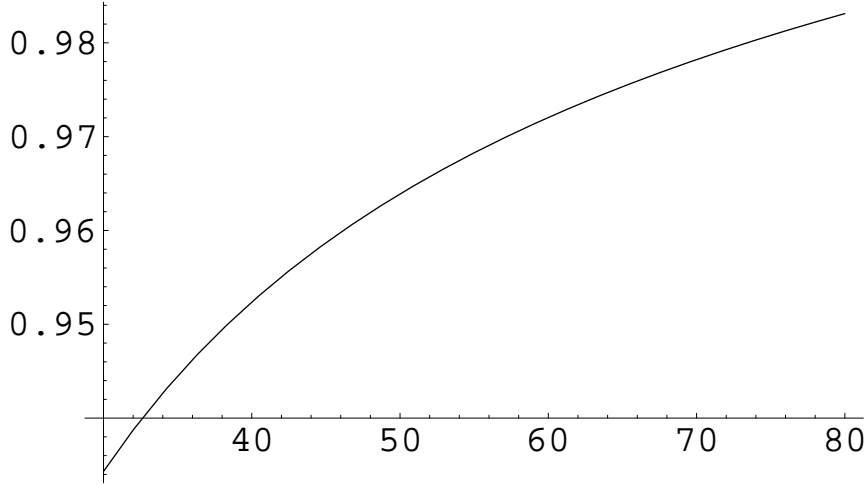}}
\caption{
The spectral index for the same parameters in Fig.3.
A horizontal axis represents the number of e-folds $N$ and
 a vertical axis represents the spectral index $n_s$.
 }
\end{figure}

So far we considered the case $\fpm$ has only the lowest-order
term $\fpm=\cpm\sigma^2/2$.  One can also incorporate even
higher-order terms in $\fpm$ such as $\fpm=c_{n\pm}(\sigma^2/2)^n$
with $n\geq 2$.  These terms can also help to realize flatter
potential.  At the same time their coefficients $c_{n\pm}$
are also constrained from the requirement that the potential
should be monotonic.  For example, if we take $n=2$, $c_{2+}=C$, and
$c_{2-}=0$, $C$ should satisfy
\beq
  C \leq \frac{11+\sqrt{125}}{2} \simeq 11.
\eeq

\section{Higher-order terms of the inflaton superfield}

Here we study the case the K\"{a}hler potential for the
inflaton superfield $S$ has a non-minimal structure.
Specifically we study the case $K$ is given by a well behaved
function $h(|S|^2)$ as
\beq
  K=h(|S|^2)+|\phip|^2+|\phim|^2,
\eeq
with the same superpotential $W=\lambda S\phip\phim$.
The scalar Lagrangian reads
\begin{eqnarray}
{\cal L}&=& -(h'+h''|S|^2)|\partial S|^2-|\partial\phip|^2
-|\partial\phim|^2-V_F-V_D, \\
V_F &=& \lambda^2 e^K
\lkk \frac{(1+h'|S|^2)^2}{h'+h''|S|^2}|\phip\phim|^2
+|S\phip|^2+|S\phim|^2+|S\phip\phim|^2+
|S|^2(|\phip|^4+|\phim|^4)\rkk, \nonumber \\
V_D &=& \frac{g^2}{2}(q_+|\phip|^2+q_-|\phim|^2+\xi)^2, \nonumber
\end{eqnarray}
where
\beq
  h'=\left.\frac{d h(x)}{dx}\right|_{x=|S|^2},~~~
  h''=\left.\frac{d^2 h(x)}{dx^2}\right|_{x=|S|^2}.
\eeq
We consider the case $S$ serves as the inflaton and $\phi_\pm$
settle down to zero during hybrid inflation.
As before let us identify the real part of $S$,
$\sigma=\sqrt{2}{\rm Re}S$ with the inflaton.
The equation of motion for its homogeneous part reads
\beq
  \lmk h' + \frac{\sigma^2}{2}h''\rmk\ddot\sigma
+2\lmk \sigma h''+\frac{\sigma^3}{4}h'''\rmk\dot{\sigma}^2
+3H\lmk h' + \frac{\sigma^2}{2}h''\rmk\dot\sigma
+V'_{1-loop}[\sigma]=0.  \label{eqm}
\eeq
Here $V'_{1-loop}[\sigma]$ is the derivative of the scalar potential
after one-loop correction given in Eq. (\ref{Potential}).
During slow-roll inflation the first two terms in Eq. (\ref{eqm})
are negligible.  The amplitude of quantum fluctuation acquired
during the Hubble time, $\delta\sigma$ is given by
\beq
  \delta\sigma=\lmk h'+\frac{\sigma^2}{2}h''\rmk^{-1/2}\frac{H}{2\pi},
\label{qu}
\eeq
due to the non-canonical kinetic term of $\sigma$.

From Eqs. (\ref{eqm}) and (\ref{qu}) we find that the amplitude of
comoving curvature perturbation is given by
\beq
\mathcal{P}^{1/2}_{\zeta}
=\frac{3H^3}{2\pi |V'_{1-loop}[\sigma]|}
\lmk h'+\frac{\sigma^2}{2}h''\rmk^{1/2}.
\eeq
Thus one could obtain a larger amplitude of perturbation with the
same $\xi$ is the factor $h'+\frac{\sigma^2}{2}h''$ takes a value
larger than unity and this may also help to reduce $\xi$.

Finally, we consider the case of the most general K\"{a}hler potential
\beq
  K=h(|S|^2)+|\phip|^2+|\phim|^2+f_+(|S|^2)|\phip|^2+f_-(|S|^2)|\phim|^2.
\eeq
Eqs.(\ref{mphi}) and (\ref{mfer}) are replaced with
\begin{eqnarray}
&& m_{\varphi_{\pm}}^2 = q_{\pm}g^2\xi+e^{h(|S|^2)}
\frac{\lambda^2|S|^2}{(1+\fp)(1+\fm)} \\
&& m_{fermion}^2 = e^{h(|S|^2)}\frac{\lambda^2|S|^2}{(1+\fp)(1+\fm)}.
\end{eqnarray}
Repeating similar calculations, we obtain
\begin{eqnarray}
\mathcal{P}^{1/2}_{\zeta} = \frac{4\pi\xi}{\sqrt{6}g}
\lmk \frac{2}{\sigma}+\sigma h' -\frac{\fps}{1+\fp}-\frac{\fms}{1+\fm} \rmk^{-1}
\lmk h'+\frac{\sigma^2}{2}h''\rmk^{1/2},
\end{eqnarray}
which shows the above mentioned two features, namely, the enhancement
 by the mixing terms $f_{\pm}$ and the canonical normalization of
the inflaton.  Thus these effects can cooperate with each other to
further reduce the constraint on $\xi$. 

For a specific example,
if $h(|S|^2)$ contain higher-order terms such as $|S|^{(2+2n)}$,
$(n=1,2,...)$, in addition to the canonical part $|S|^2$,
 they will help to allow a lower value of $\xi$
as long as their coefficients are positive.  If some of the
higher-order term had a negative contribution, the resultant
amplitude of perturbation might be somewhat lowered.  Such cases
with negative coefficients may well result in
positive non definite kinetic term causing instability.
Hence it is unlikely that these cases are phenomenologically viable.

\section{Conclusion}

In this paper, we have shown that the higher-order terms in
 the K\"{a}hler potential in general affect the dynamics of
hybrid inflation and
 the generated density perturbation significantly.
Although these terms do not lead the $\eta$ problem,
 they alter the slope of the potential and the dynamics of the inflaton.
This means that it is indispensable
 for a quantitative analysis of D-term inflation to take account of these terms.

Note that the modification is caused by the higher term
 with a coefficient $c_+$ of the order of unity.
It means that these higher terms become more significant
 as the cut-off scale of higher dimensional operators decreases,
 because the coefficients of these term $c_{\pm}$ in the
K\"{a}hler potential
 is replaced by $c_{\pm}/M^2$ with a cut-off scale $M \ll 1$.
For example, this situation can be realized
 in Type I or Type IIB orientifold models \cite{Halyo, Dasgupta},
 since the string scale is not necessarily the Planck scale in
these theories.
As another noteworthy fact, in the case that the leading term
for the inflaton
 in the K\"{a}hler potential $|S|^2$ is replaced with $|S|^2/M^2$,
the slow roll parameter $\eta$ is given as
\begin{equation}
\eta \simeq \frac{g^2}{8\pi^2}\frac{1}{M^2}
 = 1 \left(\frac{g^2}{0.8}\right)\left(\frac{10^{-1}}{M}\right)^2,
\end{equation}
hence we find that the D-term inflation is also faced with $\eta$ problem
 by a supergravity effect and is impossible for such a low cut-off model.

The higher-order terms in the K\"{a}hler potential can make the potential
 for the inflaton flatter than logarithmic.
The flat potential enables the reduced FI term to accomplish
the generation of
 the appropriate density perturbation
 and yields the different constraint on the magnitude of the FI term.
Then,
 the influence of the cosmic string formed after inflation on
CMB spectrum
can be suppressed to an acceptable level.

Thus, we can conclude that the D-term inflation model can be consistent
with the absence of CMB signature from cosmic strings,
 even if the Yukawa coupling $\lambda$ is not extremely small.
Remarkably, our proposal predicts the existence of cosmic strings
 unlike other solutions to the cosmic string problem in the
literature where
 the model was modified so as not to
form cosmic strings \cite{Urrestilla, Dasgupta}.
Since cosmic strings with $G\mu =\mathcal{O}(10^{-7})$ in our model
 is detectable, our model is testable by observations in near future.
Although the model with a very small Yukawa (and a gauge, if acceptable,)
coupling also predicts
 the existence of cosmic strings, these models are distinguishable
 because they predict different spectral indexes.
While the spectral index of the model with a very small Yukawa coupling
 \cite{Rocher, Endo} is estimated as
\begin{eqnarray}
n_s -1 \simeq -\frac{\lambda^2}{4\pi^2\xi}
=-0.0003\left(\frac{\lambda}{10^{-4}}\right)^2\left(\frac{10^{-6}}{\xi}\right),
\end{eqnarray}
that of our model is $n_s \sim 0.97$ as in Figs. 2 and 4.

%
\section*{Acknowledgements}
This work was partially supported by PPARC(OS) and by the JSPS
  Grant-in-Aid for Scientific Research No.\ 16340076(JY).
  



\begin{thebibliography}{99}

\bibitem{Inflation}
A.~H.~Guth, Phys.\ Rev.\ D {\bf 23}, 347 (1981); \\
K.~Sato, Mon.\ Not.\ Roy.\ Astron.\ Soc.\  {\bf 195}, 467 (1981); \\
for a review of inflation, see, e.g.\ A.D.\ Linde, Particle Physics and
Inflationary Cosmology (Harwood, Chur, Switzerland, 1990).

\bibitem{DtermInflation}
E.~Halyo, Phys.\ Lett.\ B {\bf 387}, 43 (1996); \\
P.~Binetury and G.~R.~Dvali, Phys.\ Lett.\ B {\bf 388}, 241 (1996).

\bibitem{LythRiotto}
D.~H.~Lyth and A.~Riotto, Phys.\ Rept.\  {\bf 314}, 1 (1999).

\bibitem{Jeannerot}
R.~Jeannerot, Phys.\ Rev.\ D {\bf 56}, 6205 (1997).

\bibitem{Hybrid}
A.~D.~Linde, Phys.\ Rev.\ D {\bf 49}, 748 (1994).

\bibitem{Kolda}
C.~Kolda and J.~March-Russell, Phys.\ Rev.\ D {\bf 60}, 023504 (1999).

\bibitem{King}
S.~F.~King and A.~Riotto, Phys.\ Lett.\ B {\bf 442}, 68 (1998) ; \\
T.~Higaki, T.~Kobayashi and O.~Seto, JHEP {\bf 0407}, 035 (2004).

\bibitem{Rocher}
J.~Rocher and M.~Sakellariadou, Phys.\ Rev.\ Lett. {\bf 94}, 011303 (2005).

\bibitem{Endo}
M.~Endo, M.~Kawasaki and T.~Moroi, Phys.\ Lett.\ B {\bf 569}, 73 (2003).

\bibitem{Urrestilla}
J.~Urrestilla, A.~Achucarro and A.~C.~Davis,
 Phys.\ Rev.\ Lett.\  {\bf 92}, 251302 (2004).
 
\bibitem{Other}
M.~Wyman, L.~Pogosian and I.~Wasserman, Phys.\ Rev.\ D {\bf 72}, 023513 (2005); \\
A.~A.~Fraisse, astro-ph/0503402.

\bibitem{Linde}
We thank Andrei Linde for telling us this possibility.

\bibitem{Hilltop}
L.~Boubekeur and D.~H.~Lyth, JCAP {\bf 0507}, 010 (2005).

\bibitem{Halyo}
E.~Halyo, Phys.\ Lett.\ B {\bf 454}, 223 (1999); \\
T.~Kobayashi and O.~Seto, Phys.\ Rev.\ D {\bf 69}, 023510 (2004).

\bibitem{Dasgupta}
K.~Dasgupta, J.~P.~Hsu, R.~Kallosh, A.~Linde and M.~Zagermann,
 JHEP {\bf 0408}, 030 (2004).


\end{thebibliography}
\end{document}